\documentclass[12pt]{article}
 \usepackage{epsfig}
 \usepackage{pstricks}
 \usepackage{pst-coil}

\input{def.cmd}

\catcode`@=11
\def\@citex[#1]#2{\if@filesw\immediate\write\@auxout{\string\citation{#2}}\fi
  \def\@citea{}\@cite{\@for\@citeb:=#2\do
    {\@citea\def\@citea{,\penalty\@m}\@ifundefined
      {b@\@citeb}{{\bf ?}\@warning
       {Citation `\@citeb' on page \thepage \space undefined}}%
\hbox{\csname b@\@citeb\endcsname}}}{#1}}

\def\citer{\@ifnextchar [{\@tempswatrue\@citexr}{\@tempswafalse\@citexr[]}}
\def\@citexr[#1]#2{\scriptsize 
  \if@filesw\immediate\write\@auxout{\string\citation{#2}}\fi
  \def\@citea{}\@cite{\@for\@citeb:=#2\do
    {\@citea\def\@citea{-\penalty\@m}\@ifundefined
       {b@\@citeb}{{\bf ?}\@warning
       {Citation `\@citeb' on page \thepage \space undefined}}%
\hbox{\csname b@\@citeb\endcsname}}}{#1}\normalsize}
\catcode`@=12

\catcode`\@=11
\long\def\@makefntext#1{
\protect\noindent \hbox to 3.2pt {\hskip-.9pt  
$^{{\ninerm\@thefnmark}}$\hfil}#1\hfill}                

\def\@makefnmark{\hbox to 0pt{$^{\@thefnmark}$\hss}}  
        
\def\ps@myheadings{\let\@mkboth\@gobbletwo
\def\@oddhead{\hbox{}
\rightmark\hfil\ninerm\thepage}   
\def\@oddfoot{}\def\@evenhead{\ninerm\thepage\hfil
\leftmark\hbox{}}\def\@evenfoot{}
\def\sectionmark##1{}\def\subsectionmark##1{}}

\newcounter{sectionc}\newcounter{subsectionc}\newcounter{subsubsectionc}
\renewcommand{\section}[1] {\vspace*{0.6cm}\addtocounter{sectionc}{1} 
\setcounter{subsectionc}{0}\setcounter{subsubsectionc}{0}\noindent 
        {\normalsize\bf\thesectionc. #1}\par\vspace*{0.4cm}}
\renewcommand{\subsection}[1] {\vspace*{0.6cm}\addtocounter{subsectionc}{1} 
        \setcounter{subsubsectionc}{0}\noindent 
        {\normalsize\it\thesectionc.\thesubsectionc. #1}\par\vspace*{0.4cm}}
\renewcommand{\subsubsection}[1]
{\vspace*{0.6cm}\addtocounter{subsubsectionc}{1}
        \noindent {\normalsize\rm\thesectionc.\thesubsectionc.\thesubsubsectionc. 
        #1}\par\vspace*{0.4cm}}

\newcounter{appendixc}
\newcounter{subappendixc}[appendixc]
\newcounter{subsubappendixc}[subappendixc]

\renewcommand{\appendix}[1] {\vspace*{0.6cm}
        \refstepcounter{appendixc}
        \setcounter{figure}{0}
        \setcounter{table}{0}
        \setcounter{equation}{0}
        \renewcommand{\thefigure}{\Alph{appendixc}.\arabic{figure}}
        \renewcommand{\thetable}{\Alph{appendixc}.\arabic{table}}
        \renewcommand{\theappendixc}{\Alph{appendixc}}
        \renewcommand{\theequation}{\Alph{appendixc}.\arabic{equation}}
        \noindent{\bf Appendix \theappendixc #1}\par\vspace*{0.4cm}}

\renewenvironment{thebibliography}[1]
        {\begin{list}{\arabic{enumi}.}
        {\usecounter{enumi}\setlength{\parsep}{0pt}
\setlength{\leftmargin 1.25cm}{\rightmargin 0pt}
         \setlength{\itemsep}{0pt} \settowidth
        {\labelwidth}{#1.}\sloppy}}{\end{list}}

\newcounter{itemlistc}
\newcounter{romanlistc}
\newcounter{alphlistc}
\newcounter{arabiclistc}

\newcommand{\fcaption}[1]{
        \refstepcounter{figure}
        \setbox\@tempboxa = \hbox{\footnotesize Fig.~\thefigure. #1}
        \ifdim \wd\@tempboxa > 6in
           {\begin{center}
        \parbox{6in}{\footnotesize\baselineskip=12pt Fig.~\thefigure. #1}
            \end{center}}
        \else
             {\begin{center}
             {\footnotesize Fig.~\thefigure. #1}
              \end{center}}
        \fi}

\newcommand{\tcaption}[1]{
        \refstepcounter{table}
        \setbox\@tempboxa = \hbox{\footnotesize Table~\thetable. #1}
        \ifdim \wd\@tempboxa > 6in
           {\begin{center}
        \parbox{6in}{\footnotesize\baselineskip=12pt Table~\thetable. #1}
            \end{center}}
        \else
             {\begin{center}
             {\footnotesize Table~\thetable. #1}
              \end{center}}
        \fi}

\def\@citex[#1]#2{\if@filesw\immediate\write\@auxout
        {\string\citation{#2}}\fi
\def\@citea{}\@cite{\@for\@citeb:=#2\do
        {\@citea\def\@citea{,}\@ifundefined
        {b@\@citeb}{{\bf ?}\@warning
        {Citation `\@citeb' on page \thepage \space undefined}}
        {\csname b@\@citeb\endcsname}}}{#1}}

\newif\if@cghi
\def\cite{\@cghitrue\@ifnextchar [{\@tempswatrue
        \@citex}{\@tempswafalse\@citex[]}}
\def\citelow{\@cghifalse\@ifnextchar [{\@tempswatrue
        \@citex}{\@tempswafalse\@citex[]}}
\def\@cite#1#2{{$\null^{#1}$\if@tempswa\typeout
        {IJCGA warning: optional citation argument 
        ignored: `#2'} \fi}}

 1
 1
 1

\font\ninerm=cmr9

\begin{document}

\date{}
\title{
{\large\rm DESY 98-129}\hfill{\large\tt hep-ph/9809265}\vspace*{3cm}\\
{\bf Baryon Asymmetry, Lepton Mixing\\ and SO(10) Unification}
\thanks{presented at the Ringberg
 Euroconference "New Trends in Neutrino Physics", Ringberg Castle, 
 Germany, May 1998}
}
\author{M. Pl\"umacher\\
{\normalsize\it Deutsches Elektronen-Synchrotron DESY, 22603 Hamburg, Germany}
\vspace*{1cm}\\                     
}        

\maketitle

\thispagestyle{empty}

\begin{abstract}
\noindent
Baryogenesis appears to require lepton number violation. This is
naturally realized in extensions of the standard model containing
right-handed neutrinos. We discuss the generation of a baryon
asymmetry by the out-of-equilibrium decay of heavy Majorana neutrinos
in these models, without and with supersymmetry. All relevant lepton
number violating scattering processes which can inhibit the generation
of an asymmetry are taken into account. We assume a similar pattern of
mixings and masses for neutrinos and up-type quarks, as suggested by
SO(10) unification. This implies that $B-L$ is broken at the
unification scale $\Lambda_{\mbox{\tiny GUT}}\sim 10^{16}\;$GeV,
if $m_{\n_\m} \sim 3\cdot10^{-3}\;$eV, as preferred by the MSW
solution to the solar neutrino deficit. The observed baryon asymmetry
is then obtained without any fine tuning of parameters.

\end{abstract}

\newpage

\section{Standard model with right-handed neutrinos}

  Baryon number ($B$) and lepton number ($L$) are not conserved in the
  standard model. At temperatures above the critical temperature of
  the electroweak phase transition $(B+L)$ violating sphaleron
  processes are in thermal equilibrium\cite{sphal}. Hence, the
  cosmological baryon asymmetry appears to require $B-L$
  violation, and therefore $L$ violation. Lepton number violation is
  naturally realized by adding right-handed Majorana neutrinos to the
  standard model.  Heavy right-handed Majorana neutrinos, whose
  existence is predicted by theories based on gauge groups containing
  SO(10)\cite{so10}, can also explain the smallness of the light
  neutrino masses via the see-saw mechanism\cite{seesaw}.

  The most general Lagrangian for couplings and masses of charged
  leptons and neutrinos is given by 
  \beq
    \cl_Y = \overline{l_{\mbox{\tiny L}}}\,H\,\l^*_l\,e_{\mbox{\tiny R}}
      +\overline{l_{\mbox{\tiny L}}}\e H^{\dg}\,\l_{\n}^*\,\n_{\mbox{\tiny R}}
      -{1\over2}\,\overline{\n_{\mbox{\tiny R}}^c}\,M\,\n_{\mbox{\tiny R}}
      +\mbox{ h.c.}\;,
  \label{yuk}
  \eeq
  The vacuum expectation value of the Higgs field $\VEV{H^0}=v\ne0$
  generates Dirac masses $m_l=\l_l\,v$ and $m_{\scr D}=\l_{\n}\,v$ for
  charged leptons and neutrinos, which are assumed to be much smaller
  than the Majorana masses $M$.
  
  \begin{figure}[t]
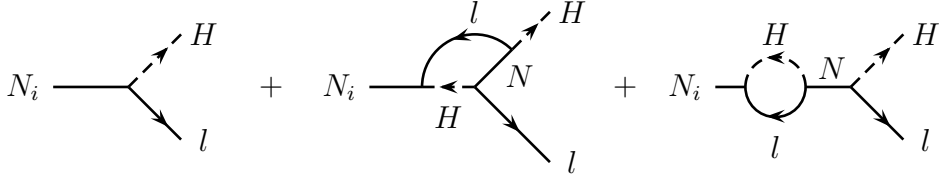

    \begin{center}
\psset{unit=1cm}
\pspicture(0,0.5)(3.7,2.1)
\psline[linewidth=1pt](0.6,1.3)(1.6,1.3)
\psline[linewidth=1pt](1.6,1.3)(2.3,0.6)
\psline[linewidth=1pt,linestyle=dashed](1.6,1.3)(2.3,2.0)
\psline[linewidth=1pt]{->}(2.03,0.87)(2.13,0.77)
\psline[linewidth=1pt]{->}(2.03,1.73)(2.13,1.83)
\rput[cc]{0}(0.2,1.3){$N_i$}
\rput[cc]{0}(2.6,0.6){$l$}
\rput[cc]{0}(2.6,2.0){$H$}
\rput[cc]{0}(3.5,1.3){$+$}
\endpspicture
\pspicture(-0.5,0.5)(4.2,2.1)
\psline[linewidth=1pt](0.6,1.3)(1.3,1.3)
\psline[linewidth=1pt,linestyle=dashed](1.3,1.3)(2.0,1.3)
\psline[linewidth=1pt](2,1.3)(2.5,1.8)
\psline[linewidth=1pt,linestyle=dashed](2.5,1.8)(3,2.3)
\psline[linewidth=1pt](2,1.3)(3,0.3)
\psarc[linewidth=1pt](2,1.3){0.7}{45}{180}
\psline[linewidth=1pt]{<-}(1.53,1.3)(1.63,1.3)
\psline[linewidth=1pt]{<-}(1.7,1.93)(1.8,1.96)
\psline[linewidth=1pt]{->}(2.75,2.05)(2.85,2.15)
\psline[linewidth=1pt]{->}(2.5,0.8)(2.6,0.7)
\rput[cc]{0}(0.2,1.3){$N_i$}
\rput[cc]{0}(1.65,0.9){$H$}
\rput[cc]{0}(2,2.3){$l$}
\rput[cc]{0}(2.6,1.45){$N$}
\rput[cc]{0}(3.3,2.3){$H$}
\rput[cc]{0}(3.3,0.3){$l$}
\rput[cc]{0}(4.0,1.3){$+$}
\endpspicture
\pspicture(-0.5,0.5)(3.5,2.1)
\psline[linewidth=1pt](0.5,1.3)(0.9,1.3)
\psline[linewidth=1pt](1.7,1.3)(2.3,1.3)
\psarc[linewidth=1pt](1.3,1.3){0.4}{-180}{0}
\psarc[linewidth=1pt,linestyle=dashed](1.3,1.3){0.4}{0}{180}
\psline[linewidth=1pt]{<-}(1.18,1.69)(1.28,1.69)
\psline[linewidth=1pt]{<-}(1.18,0.91)(1.28,0.91)
\psline[linewidth=1pt](2.3,1.3)(3.0,0.6)
\psline[linewidth=1pt,linestyle=dashed](2.3,1.3)(3.0,2.0)
\psline[linewidth=1pt]{->}(2.73,0.87)(2.83,0.77)
\psline[linewidth=1pt]{->}(2.73,1.73)(2.83,1.83)
\rput[cc]{0}(1.3,0.5){$l$}
\rput[cc]{0}(1.3,2){$H$}
\rput[cc]{0}(2.05,1.55){$N$}
\rput[cc]{0}(0.1,1.3){$N_i$}
\rput[cc]{0}(3.3,0.6){$l$}
\rput[cc]{0}(3.3,2.0){$H$}
\endpspicture
\end{center}
  \caption{Contributions to the decay of a heavy Majorana neutrino
           \label{decay_fig} }
  \end{figure}
  
  Out-of-equilibrium decays of right-handed neutrinos can generate a
  lepton asymmetry, which is then partially transformed into a baryon
  asymmetry by sphaleron processes\cite{fy}. The decay width of $N_i$
  in its rest frame reads at tree level,
  \beq
    \G_{Di}=\G_{rs}\left(N^i\to l H \right)+
    \G_{rs}\left(N^i\to \bar{l} H^{\dg} \right)
    ={(m_{\scr D}^{\dg}m_{\scr D})_{ii}\over 8\p v^2}M_i\;.
    \label{decay}
  \eeq
  Interference between the tree-level amplitude and one-loop
  corrections (cf.~fig.~\ref{decay_fig}) gives rise to a $CP$ 
  asymmetry\cite{bp_new}
  \beq
    \ve_i=
      {\Gamma(N_i\rightarrow l H)-\Gamma(N_i\rightarrow \bar{l} H^{\dg})\over
      \Gamma(N_i\rightarrow l H)+\Gamma(N_i\rightarrow \bar{l} H^{\dg})}
      ={-1\over8\pi v^2}\;\sum\limits_{j\ne i}
      {\mbox{Im}\left[\mmij^2\right]\over\mmii}\,f
      \left({M_j^2\over M_i^2}\right)\label{cpasymm}\;,
  \eeq
  where
  \beq
    f(x)=\sqrt{x}\left[\,{2-x\over 1-x}
    -(1+x)\ln\left({1+x\over x}\right)\right]\;.
  \eeq

  \begin{figure}[b]
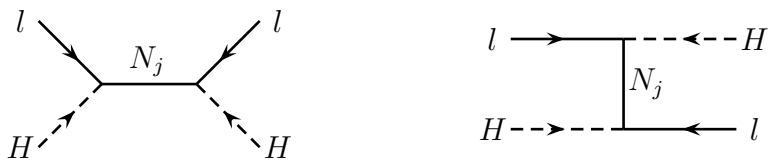

     \begin{center}
\psset{unit=6mm}
  \pspicture[0.5](1,0.7)(8,3.3)
    \psline[linewidth=1pt,linestyle=dashed](1.6,0.6)(3,2)
    \psline[linewidth=1pt]{->}(2.3,1.3)(2.4,1.4)
    \rput[cc]{0}(1.2,0.6){$H$}
    \psline[linewidth=1pt](1.6,3.4)(3,2)
    \psline[linewidth=1pt]{->}(2.4,2.6)(2.5,2.5)
    \rput[cc]{0}(1.2,3.4){$l$}
    \psline[linewidth=1pt](3,2)(5.1,2)
    \rput[cc]{0}(4,2.5){$N_j$}
    \psline[linewidth=1pt](5.1,2)(6.5,3.4)
    \psline[linewidth=1pt]{->}(5.7,2.6)(5.6,2.5)
    \rput[cc]{0}(6.9,3.4){$l$}
    \psline[linewidth=1pt,linestyle=dashed](5.1,2)(6.5,0.6)
    \psline[linewidth=1pt]{->}(5.8,1.3)(5.7,1.4)
    \rput[cc]{0}(6.9,0.6){$H$}
  \endpspicture
  \hspace{2cm}
  \pspicture[0.5](8.5,0.7)(15,3.3)
    \psline[linewidth=1pt](9,3)(11.5,3)
    \psline[linewidth=1pt]{->}(10.1,3)(10.2,3)
    \rput[cc]{0}(8.6,3){$l$}
    \psline[linewidth=1pt,linestyle=dashed](11.5,3)(14,3)
    \psline[linewidth=1pt]{->}(12.9,3)(12.8,3)
    \rput[cc]{0}(14.4,3){$H$}
    \psline[linewidth=1pt](11.5,3)(11.5,1)
    \rput[cc]{0}(12,2){$N_j$}
    \psline[linewidth=1pt,linestyle=dashed](9,1)(11.5,1)
    \psline[linewidth=1pt]{->}(10.1,1)(10.2,1)
    \rput[cc]{0}(8.6,1){$H$}
    \psline[linewidth=1pt](11.5,1)(14,1)
    \psline[linewidth=1pt]{->}(12.9,1)(12.8,1)
    \rput[cc]{0}(14.4,1){$l$}
  \endpspicture
\end{center}
     \caption{\it Lepton number violating lepton Higgs scattering
       \label{lept_fig}}
  \end{figure}
  \begin{figure}[t]
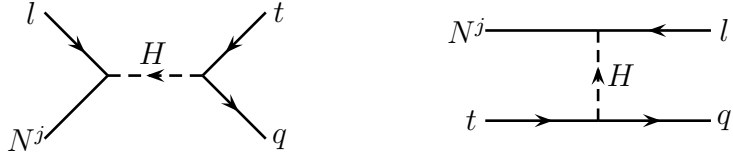

      \begin{center}
  \psset{unit=6mm}
  \pspicture[0.5](1.0,0.7)(7.3,3.3)
    \psline[linewidth=1pt](1.6,0.6)(3,2)
    \rput[cc]{0}(1.2,0.6){$N^j$}
    \psline[linewidth=1pt](1.6,3.4)(3,2)
    \psline[linewidth=1pt]{->}(2.4,2.6)(2.5,2.5)
    \rput[cc]{0}(1.3,3.4){$l$}
    \psline[linewidth=1pt,linestyle=dashed](3,2)(5.1,2)
    \psline[linewidth=1pt]{->}(3.95,2)(3.85,2)
    \rput[cc]{0}(4,2.5){$H$}
    \psline[linewidth=1pt](5.1,2)(6.5,3.4)
    \psline[linewidth=1pt]{->}(5.7,2.6)(5.6,2.5)
    \rput[cc]{0}(6.8,3.4){$t$}
    \psline[linewidth=1pt](5.1,2)(6.5,0.6)
    \psline[linewidth=1pt]{->}(5.8,1.3)(5.9,1.2)
    \rput[cc]{0}(6.8,0.6){$q$}
  \endpspicture
  \hspace{2cm}
  \pspicture[0.5](8.5,0.7)(14.5,3.3)
    \psline[linewidth=1pt](9,3)(11.5,3)
    \rput[cc]{0}(8.6,3){$ N^j$}
    \psline[linewidth=1pt](11.5,3)(14,3)
    \psline[linewidth=1pt]{<-}(12.6,3)(12.7,3)
    \rput[cc]{0}(14.3,3){$ l$}
    \psline[linewidth=1pt,linestyle=dashed](11.5,3)(11.5,1)
    \psline[linewidth=1pt]{->}(11.5,2.1)(11.5,2.2)
    \rput[cc]{0}(12,2){$H$}
    \psline[linewidth=1pt](9,1)(11.5,1)
    \psline[linewidth=1pt]{->}(10.3,1)(10.5,1)
    \rput[cc]{0}(8.7,1){$ t$}
    \psline[linewidth=1pt](11.5,1)(14,1)
    \psline[linewidth=1pt]{->}(12.8,1)(12.9,1)
    \rput[cc]{0}(14.3,1){$ q$}
  \endpspicture
\end{center}
      \caption{\it Lepton number violating neutrino top-quark scattering
       \label{top_fig}} 
  \end{figure}
  
  Consider now the various processes which are relevant in the
  leptogenesis scenario. In order to obtain a lepton asymmetry of the
  correct order of magnitude, the right-handed neutrinos have to be
  numerous before decaying, i.e., they have to be in thermal
  equilibrium at high temperatures. The Yukawa interactions
  (\ref{yuk}) are too weak to achieve this and additional interactions
  are therefore needed.  Since right-handed neutrinos are a necessary
  ingredient of SO(10) unified theories, it is natural to consider
  leptogenesis within an extended gauge model, contained in a SO(10)
  GUT. The minimal extension of the standard model is based on the
  gauge group
  \beq
    G=\mbox{SU}(3)_C\times\mbox{SU}(2)_{\mbox{\tiny L}}\times
     \mbox{U}(1)_Y\times\mbox{U}(1)_{Y'}\ \subset\ \mbox{SO}(10)\;.
  \eeq
  Here $U(1)_{Y'}$, and therefore $B-L$, is spontaneously broken, and
  the breaking scale is related to the heavy neutrino masses.  The
  additional neutral gauge boson $Z'$ accounts for pair creation and
  annihilation processes and for flavour transitions between heavy
  neutrinos of different generations. For appropriately chosen
  parameters these processes generate an equilibrium distribution of
  heavy neutrinos at high temperatures\cite{pluemi}.
  
  Of crucial importance are the $\Delta L=2$ lepton number violating
  scatterings shown in fig.~\ref{lept_fig} which, if too strong, erase
  any lepton asymmetry. Similarly, the $\Delta L=1$ lepton number
  violating neutrino top-quark scatterings shown in fig.~\ref{top_fig}
  have to be taken into account because of the large top Yukawa
  coupling. Finally, the heavy neutrino decays
  (cf.~fig.~\ref{decay_fig}) as well as the inverse decays have to be
  incorporated in the Boltzmann equations.
  
  Based on these equations the resulting lepton and baryon asymmetries
  can be evaluated\cite{luty,pluemi}, and one may
  ask whether the right order of magnitude of the asymmetry results
  naturally in the leptogenesis scenario. To address this question one
  has to discuss patterns of neutrino mass matrices which determine
  the generated asymmetry.

\section{Neutrino masses and mixings}
  In the following we shall assume a similar pattern of mixings and
  mass ratios for leptons and quarks\cite{bp}, which is natural in
  SO(10) unification.  Such an ansatz is most transparent in a basis
  where all mass matrices are maximally diagonal. In addition to real
  mass eigenvalues two mixing matrices then appear. One can always
  choose a basis for the lepton fields such that the mass matrices
  $m_l$ for the charged leptons and $M$ for the heavy Majorana
  neutrinos $N_i$ are diagonal with real and positive eigenvalues.  In
  this basis $m_{\scr D}$ is a general complex matrix, which can be
  diagonalized by a biunitary transformation. Therefore, we can write
  $m_{\scr D}$ as product of a diagonal matrix $m_{\scr D,diag}$ and
  two unitary matrices $V$ and $U^{\dag},$
  \beq
    m_{\scr D}=V\,m_{\scr D,diag}\,U^{\dag}\;,
  \eeq
  where the eigenvalues $m_i$ of $m_{\scr D,diag}$ are real and
  positive. In the absence of a Majorana mass term $V$ and $U$ would 
  correspond to Kobayashi-Maskawa type mixing matrices of left- and 
  right-handed charged currents, respectively.
  
  According to eq.~(\ref{cpasymm}) the $CP$ asymmetry is determined by
  the mixings and phases present in the product $m_{\scr D}^{\dg}
  m_{\scr D}$, where the matrix $V$ drops out.  Hence, to
  leading order, the mixings and phases which are responsible for
  baryogenesis are entirely determined by the matrix $U$.
  Correspondingly, the mixing matrix in the leptonic charged
  current, which determines $CP$ violation and mixings of the light
  leptons, depends on mass ratios and mixing angles and phases of $U$
  and $V$.  This implies that there exists no direct connection
  between the $CP$ violation and generation mixing relevant at high
  and low energies.

  Consider now the mixing matrix $U$. One can factor out five phases,
  \beq
    U=\mbox{e}^{i\g}\,\mbox{e}^{i\l_3\a}\,\mbox{e}^{i\l_8\b}\,U_1\,
    \mbox{e}^{i\l_3\s}\,\mbox{e}^{i\l_8\t}\;,
  \eeq  
  where the $\l_i$ are the Gell-Mann matrices. The remaining matrix
  $U_1$ depends on three mixing angles and one phase, like the CKM
  matrix for quarks. In analogy to the quark mixing matrix we choose
  the Wolfenstein parametrization\cite{wolfenstein} as ansatz for
  $U_1$,
  \beq
    U_1=\left(\begin{array}{ccc}
    \scriptstyle 1-\l^2/2  & \scriptstyle\l &\scriptstyle A\l^3(\r-i\h)\\
    \scriptstyle -\l &\scriptstyle 1-\l^2/2 &\scriptstyle  A\l^2 \\
    \scriptstyle A\l^3(1-\r-i\h) &\scriptstyle -A\l^2 &\scriptstyle 1
    \end{array}\right)\;,
    \label{mm}
  \eeq
  where $A$ and $|\r+i\h|$ are $\co(1)$, while the mixing
  parameter $\l$ is assumed to be small. For the masses $m_i$
  and the eigenvalues $M_i$ of the Majorana mass matrix $M$ we assume 
  the same hierarchy which is observed for up-type quarks,
  \beqa
    m_1=b\l^4m_3\;,&\quad m_2=c\l^2m_3\;,&\quad b,c=\co(1)\\
    M_1=B\l^4M_3\;,&\quad M_2=C\l^2M_3\;,&\quad B,C=\co(1)\;.\label{Mmass}
  \eeqa  
  For the eigenvalues $m_i$ of the Dirac mass matrix this choice is
  suggested by SO(10) unification. For the masses $M_i$ this is an
  assumption motivated by simplicity. The masses $M_i$ cannot be
  degenerate, because in this case there exists a basis for
  $\n_{\mbox{\tiny R}}$ such that $U = 1$, which implies that no
  baryon asymmetry is generated. However, the precise form of the
  assumed hierarchy has no influence on the viability of the
  leptogenesis mechanism\cite{bp}. The see-saw mechanism then yields
  light neutrino masses,
  \beqa
     m_{\n_e}&=&{b^2\over\left|C+\mbox{e}^{4i\a}\;B\right|}\;\l^4
             \;m_{\n_{\t}}+\co\left(\l^6\right)\label{mne}\;,\\
     m_{\n_{\m}}&=&{c^2\left|C+\mbox{e}^{4i\a}\;B\right|\over BC}
             \;\l^2\;m_{\n_{\t}}+\co\left(\l^4\right)\label{mnm}\;,\\
     m_{\n_{\t}}&=&{m_3^2\over M_3}+\co\left(\l^4\right)\;.\label{mnt}
  \eeqa
 
  The $CP$-asymmetry in the decay of the lightest right-handed
  neutrino $N_1$ is easily obtained from eqs.~(\ref{cpasymm}) and
  (\ref{mm})-(\ref{Mmass}),  
  \beq
    \ve_1=-\;{3\over16\p}\;{B\;A^2\over c^2+A^2\;|\r+i\h|^2}\;\l^4\;
    {m_3^2\over v^2}\;\mbox{Im}\left[(\r-i\h)^2
    \mbox{e}^{i2(\a+\sqrt{3}\b)}\right]
    \;+\;\co\left(\l^6\right)\;.\label{cpa}
  \eeq
  For $\l \sim 0.1$ one
  has $|\ve_1|\leq 10^{-6}\cdot m_3^2/v^2$. Hence, a large value of the
  Yukawa coupling $m_3/v$ will be required by this mechanism of
  baryogenesis. This holds irrespective of the neutrino mixings and the
  heavy neutrino masses.

\section{Numerical Results}

  To obtain a numerical value for the produced baryon asymmetry, one
  has to specify the free parameters in the ansatz
  (\ref{mm})-(\ref{Mmass}).  In the following we will use as a
  constraint the value for the $\n_{\m}$-mass which is preferred by
  the MSW explanation\cite{msw} of the solar neutrino deficit,
  \beq
    m_{\n_{\m}}\simeq 3\cdot10^{-3}\;\mbox{eV}\;. \label{msw}
  \eeq
  A generic choice for the free parameters is to take all $\co(1)$
  parameters equal to one and to fix $\l$ to a value which is
  of the same order as the $\l$ parameter of the quark mixing matrix,
  \beq
    A=B=C=b=c=|\r+i\h|\simeq 1\; ,\qquad \l\simeq 0.1\;. \label{p1}
  \eeq
  {}From eqs.~(\ref{mne})-(\ref{mnt}), (\ref{msw}) and (\ref{p1}) one
  now obtains,
  \beq\label{nmasses}
    m_{\n_e}\simeq 8\cdot10^{-6}\;\mbox{eV}\; , \quad
    m_{\n_{\t}}\simeq 0.15\;\mbox{eV}\; .\label{m1}
  \eeq
  Finally, a second mass scale has to be specified. In unified
  theories based on SO$(10)$ the Dirac neutrino mass $m_3$ is
  naturally equal to the top-quark mass,
  \beq\label{3t}
     m_3=m_t\simeq 174\;\mbox{GeV}\;.\label{mtop}
  \eeq
  This determines the masses of the heavy Majorana neutrinos $N_i$,
  $M_3 \simeq 2\cdot10^{14}\;$GeV and, consequently, 
  $M_1\simeq 2\cdot10^{10}\;\mbox{GeV}$ and
  $M_2\simeq 2\cdot10^{12}\;\mbox{GeV}$. {}From eq.~(\ref{cpa}) one
  obtains the $CP$ asymmetry $|\ve_1| \simeq 3\cdot10^{-6}$, where we have
  assumed maximal phases. The solution of the Boltzmann equations now
  yields the $(B-L)$ asymmetry,
  \beq
     Y_{\mbox{\tiny B-L}} \simeq 8\cdot10^{-10}\; , \label{nonsusy_res1}
  \eeq
  which is indeed the correct order of magnitude!
  
  The large mass $M_3$ of the heavy Majorana neutrino $N_3$ suggests
  that $B-L$ is already broken at the unification scale
  $\Lambda_{\mbox{\scriptsize GUT}} \sim 10^{16}$ GeV, without any
  intermediate scale of symmetry breaking. This large value of $M_3$
  is a consequence of the choice $m_3 \simeq m_t$. To test the
  sensitivity of the result for $Y_{\mbox{\tiny B-L}}$ on this
  assumption, consider as an alternative the choice $m_3 = m_b \simeq
  4.5$ GeV, with all other parameters remaining unchanged. In this
  case one has $M_3=10^{11}$ GeV and $|\ve_1| = 2\cdot10^{-9}$ for the
  mass of $N_3$ and the $CP$ asymmetry, respectively. Since the
  maximal $B-L$ asymmetry is $-\ve_1/g*$ (cf.~\cite{kw}), it is clear
  why the generated asymmetry,
  \beq
    Y_{\mbox{\tiny B-L}}\simeq 6\cdot10^{-13}\;,\label{nonsusy_res2}
  \eeq
  is too small by more than two orders of magnitude. We conclude that
  high values for both masses $m_3$ and $M_3$ are preferred, which is
  natural in SO(10) unification.

  Models for dark matter involving massive neutrinos favour a
  $\t$-neutrino mass $m_{\n_{\t}} \simeq 5\; \mbox{eV}$\cite{raf},
  which is significantly larger than the value given in (\ref{m1}).
  Such a large value for the $\t$-neutrino mass can be accommodated
  within the ansatz described in this section. However, it does not
  correspond to the simplest choice of parameters and requires some
  fine-tuning. For the mass of the heaviest Majorana neutrino one
  obtains in this case $M_3 \simeq 6\cdot 10^{12}$ GeV.

\section{Supersymmetric extension}

  Without an intermediate scale of symmetry breaking, the unification
  of gauge couplings appears to require low-energy supersymmetry.
  Supersymmetric leptogenesis has already been considered\cite{camp}
  in an approximation where lepton number violating scatterings are
  neglected which inhibit the generation of lepton number. However, a
  full analysis of the mechanism including all the relevant scattering
  processes is necessary in order to get a reliable relation between
  the input parameters and the final asymmetry\cite{pluemi2}. It turns
  out that the lepton number violating scatterings are qualitatively
  more important than in the non-supersymmetric scenario and that they
  can even account for the generation of an equilibrium distribution
  of heavy neutrinos at high temperatures.
  \begin{figure}[t]
    \input{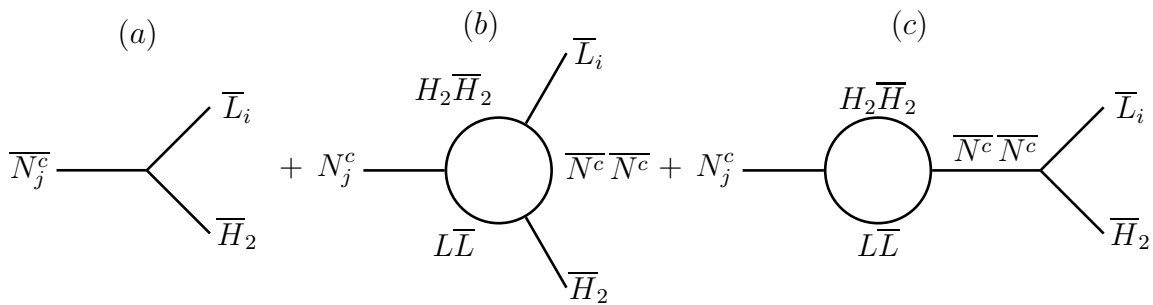}
    \caption{\it Decay modes of the right-handed Neutrino superfield.
              \label{susy_decay}}
  \end{figure}
  
  The supersymmetric generalization of the Lagrangian (\ref{yuk}) is
  the superpotential
  \beq
    W = {1\over2}N^cMN^c + \m H_1\e H_2 + H_1 \e L \l_l E^c 
        + H_2 \e L \l_{\n} N^c\;,
  \eeq
  where, in the usual notation, $H_1$, $H_2$, $L$, $E^c$ and $N^c$ are
  chiral superfields describing spin-0 and spin-${1\over 2}$ fields.
  The basis for the lepton fields can be chosen as in the
  non-supersymmetric case.  The vacuum expectation values
  $v_1=\left\langle H_1\right\rangle$ and $v_2=\left\langle
  H_2\right\rangle$ of the two neutral Higgs fields generate Dirac
  masses $m_l=\l_lv_1$ for the charged leptons and their scalar
  partners, and $m_{\scr D}=\l_{\n}v_2$ for the neutrinos.
  The heavy neutrinos and their scalar partners can decay into
  various final states which can be summarized in the superfield
  diagrams in fig.~\ref{susy_decay}. At tree level,
  the decay widths read,
  \beqa
    \G_{rs}\Big(N_i\to\wt{l}+\wt{h}^c\;\Big)
       =\G_{rs}\Big(N_i\to l+H_2\Big)
       &=&{\mmii\over 16\p v_2^2}\ M_i\;,
         \label{decay1}\\[1ex]
    \G_{rs}\Big(\sni\to\wt{l}+H_2\Big)
       =\G_{rs}\Big(\wt{N_i}\to l+\wt{h}^c\;\Big)
       &=&{\mmii\over 8\p v_2^2}\ M_i\;.
        \label{decay2}
  \eeqa
  The $CP$ asymmetry in each of the decay channels is given by
  \beqa
     \ve_i&=& {1\over8\p v_2^2}\;\sum\limits_{j\ne i}
      {\mbox{Im}\left[\mmij^2\right]\over\mmii}\;
        g\left({M_j^2\over M_i^2}\right)\\[1ex]
      &&\qquad g(x)=\sqrt{x}\;\left[\mbox{ln}\left({1+x\over x}\right)
        +{2\over x-1}\right]\;.
  \eeqa
  It arises through interference of tree level and one-loop diagrams
  shown in fig.~\ref{susy_decay}.  In the case of a mass hierarchy,
  $M_j\gg M_i$, the $CP$ asymmetry is twice as large as in the
  non-supersymmetric case.
  
  Like in the non-supersymmetric scenario lepton number violating
  scatterings mediated by a heavy (s)neutrino have to be included in a
  consistent analysis, since they can easily reduce the generated
  asymmetry by two orders of magnitude\cite{pluemi2}. A very
  interesting new feature of the supersymmetric model is that the
  (s)neutrino (s)top scatterings are strong enough to bring the
  neutrinos into thermal equilibrium at high temperatures. Hence, an
  equilibrium distribution can be reached for temperatures far below
  the masses of heavy gauge bosons.

  \begin{figure}[t]
     \begin{center}
     \epsfig{file=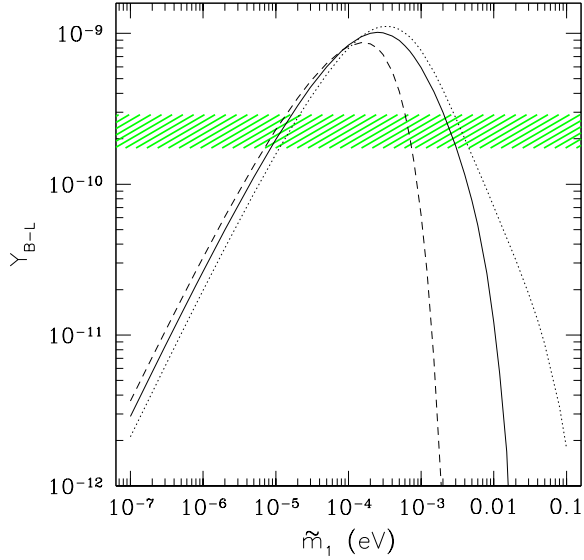,width=7.8cm}
     \end{center}
     \caption{\it The generated $(B-L)$ asymmetry for
       $M_1=10^8\;$GeV (dotted line), $M_1=10^{10}\;$GeV (solid line)
       and $M_1=10^{12}\;$GeV (dashed line). The hatched area shows
       the measured value for the asymmetry.
       \label{susy_res}}
  \end{figure}
  
  Further, it turns out that the asymmetry essentially depends on the
  ratio
  \beq
    \wt{m}_1 = {\mmoo\over M_1}\;. 
  \eeq
  For the mass matrices discussed in sect.~4 $\wt{m}_1$ is of the same
  order as the muon neutrino mass. One easily verifies,
  \beq
    \wt{m}_1 = {C(c^2+A^2|\r+i\eta|^2)\over 
                c^2 \left|C + e^{4i\alpha} B\right|}\,m_{\n_\m}\;
                + \co(\l^2)\;.
    \label{m1t}
  \eeq
  
  In fig.~\ref{susy_res} we have plotted the generated lepton
  asymmetry as function of $\wt{m}_1$ for three different values of
  $M_1$, where we have assumed the hierarchy $M_2^2=10^3\;M_1^2$,
  $M_3^2=10^6\;M_1^2$ and the $CP$ asymmetry $\ve_1=-10^{-6}$.
  
  Fig.~\ref{susy_res} demonstrates, that in the whole
  parameter range the generated asymmetry is much smaller than the
  value $\ve/g_*\sim4\cdot10^{-9}$ which one obtains 
  by neglecting lepton number violating scattering
  processes.  For small $\wt{m}_1$ the reason is that the Yukawa
  interactions are too weak to bring the neutrinos into equilibrium at
  high temperatures.  For large $\wt{m}_1$, on the other hand, the
  lepton number violating scatterings wash out a large part of the
  generated asymmetry at temperatures $T<M_1$.
  
  Baryogenesis is possible in the range
  \beq
    10^{-5}\;\mbox{eV}\;\ltap\;\wt{m}_1\;\ltap\;
    5\cdot10^{-3}\;\mbox{eV}\;.
  \eeq
  This result is independent of any assumptions on the mass matrices, 
  in particular it is not a consequence of the ansatz discussed in 
  sect.~5. This ansatz just implies $\wt{m}_1\simeq m_{\n_\m}$
  (cf.~(\ref{m1t})). It is very interesting that the $\n_{\m}$-mass
  preferred by the MSW explanation of the solar neutrino deficit lies
  indeed in the interval allowed by baryogenesis according to 
  fig.~\ref{susy_res}. 

  \begin{figure}[t]
      \mbox{ }\hspace{-0.7cm}
      \begin{minipage}[t]{7.5cm}
        \begin{center}\hspace{1cm}(a)\end{center}
        \epsfig{file=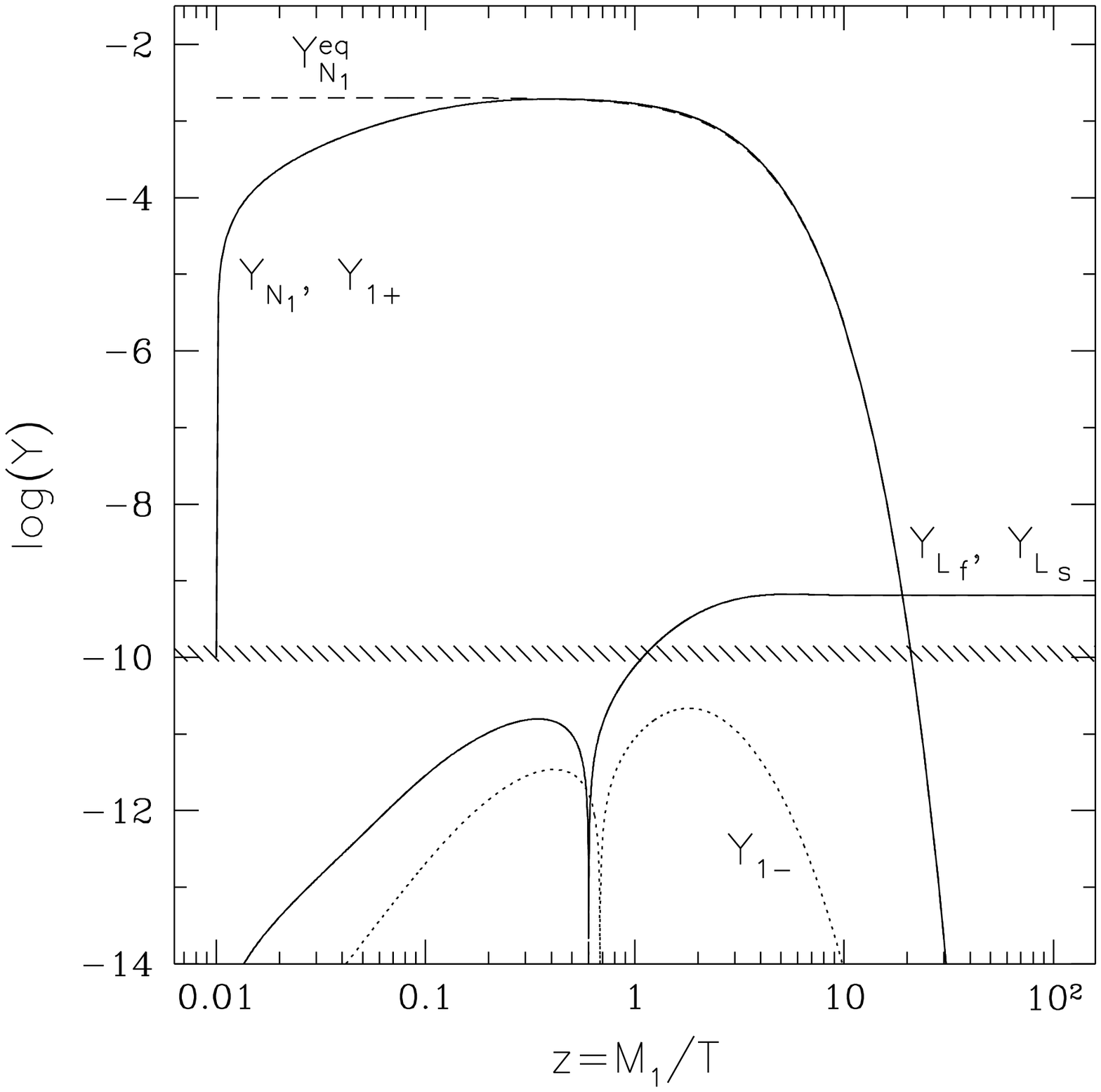,width=7.5cm}
      \end{minipage}
      \hspace{\fill}
      \begin{minipage}[t]{7.5cm}
        \begin{center}\hspace{1cm}(b)\end{center}
        \epsfig{file=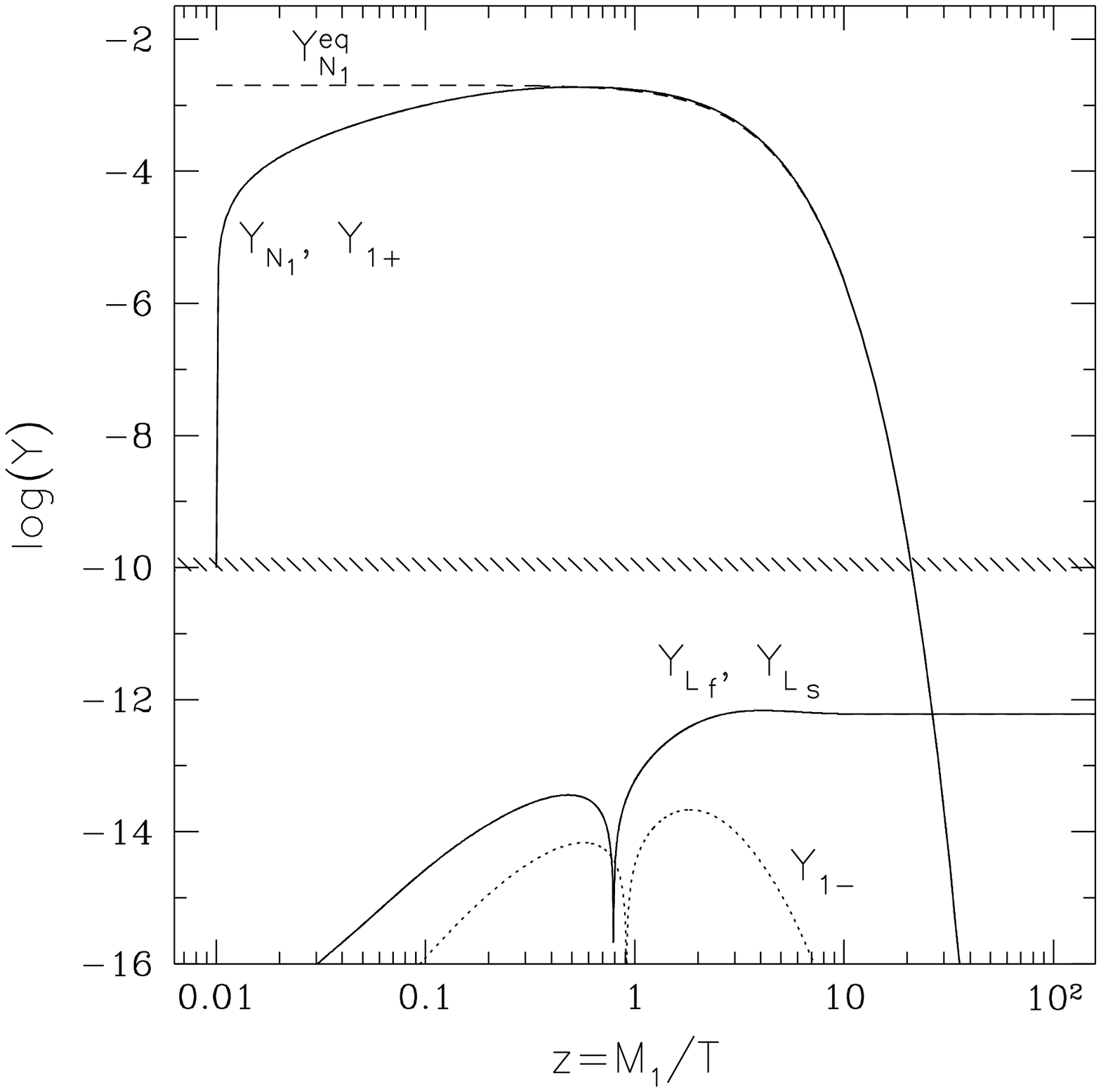,width=7.5cm}
     \end{minipage}  
       \caption{\it Generated asymmetry if one assumes a similar
        pattern of masses and mixings for the leptons and the
        quarks. In both figures we have $\l=0.1$ and $m_3=m_t$ (a) and
        $m_3=m_b$ (b). The hatched area shows the lepton asymmetry
        corresponding to the measured baryon asymmetry.
       \label{sol2fig}}
  \end{figure}
  
  Consider now again the simplest choice of parameters given by
  eqs.~({\ref{msw})-(\ref{mtop}). The corresponding generated lepton
  asymmetries are shown in fig.~\ref{sol2fig}a. $Y_{\mbox{\tiny L}_f}$
  and $Y_{\mbox{\tiny L}_s}$ denote the absolute values of the
  asymmetries stored in leptons and their scalar partners,
  respectively. They are related to the baryon asymmetry by
  \beq
    Y_{\mbox{\tiny B}} = - {8\over 23}\ Y_{\mbox{\tiny L}}\quad, 
    \qquad Y_{\mbox{\tiny L}} = Y_{\mbox{\tiny L}_f} 
    + Y_{\mbox{\tiny L}_s}\;.
  \eeq 
  $Y_{N_1}$ is the number of heavy neutrinos per comoving volume
  element, and $Y_{1\pm} = Y_{\snone}\pm Y_{\wt{N}_1}$, where 
  $Y_{\snone}$ is the number of scalar neutrinos per comoving volume
  element. As fig.~(\ref{sol2fig}a) shows, the generated baryon
  asymmetry has the correct order of magnitude,
  \beq
    Y_{\mbox{\tiny B-L}}\simeq1\cdot10^{-9}\;.\label{susy_res1}
  \eeq     
  Lowering the Dirac mass scale of the neutrinos to the bottom-quark
  scale has again dramatic consequences. The
  baryon asymmetry is reduced by three orders of magnitude
  \beq
    Y_{\mbox{\tiny B-L}}\simeq 1\cdot10^{-12}\;. \label{susy_res2}
  \eeq
  Hence, like in the non-supersymmetric scenario, large values for both
  masses $m_3$ and $M_3$ are necessary.
  
  Comparing the results (\ref{susy_res1}) and (\ref{susy_res2})
  with their non-supersymmetric counterparts
  (\ref{nonsusy_res1}) and (\ref{nonsusy_res2}),
  one sees that the larger $CP$ asymmetry and the additional
  contributions from the sneutrino decays in the supersymmetric
  scenario are compensated by the wash-out processes which are
  stronger than in the non-supersymmetric case. The final asymmetries
  are the same in the non-supersymmetric and in the supersymmetric
  case.
  
  The recently reported atmospheric neutrino anomaly\cite{tot} may be
  due to neutrino oscillations. The required mass difference and
  mixing angle are $\Delta m^2 \sim 0.005$ eV$^2$ and $\sin^2{2\Theta}
  \sim 1$. The preferred solution for baryogenesis discussed above
  yields (cf.~eq.~(\ref{nmasses})) $m_{\n_\t}^2-m_{\n_\m}^2 \simeq
  0.02$ eV$^2$ which, within the theoretical and experimental
  uncertainties, is certainly consistent with the mass difference
  required by the neutrino oscillation hypothesis. The $\n_\t$-$\n_\m$
  mixing angle is not constrained by leptogenesis and therefore a free
  parameter in principle. The large value needed, however, is against
  the spirit of small generation mixings manifest in the Wolfenstein
  ansatz and would require some special justification.

\section{Summary}

  Anomalous electroweak $B+L$ violating processes are in thermal
  equilibrium in the high-temperature phase of the standard
  model. Hence, the cosmological baryon asymmetry can be
  generated from a primordial lepton asymmetry. Necessary ingredients
  are right-handed neutrinos and Majorana masses, which occur
  naturally in SO(10) unification.
  
  The baryon asymmetry can be computed by standard methods based on
  Boltzmann equations. In a consistent analysis lepton number
  violating scatterings have to be taken into account, since they can
  erase a large part of the asymmetry. In supersymmetric scenarios
  these scatterings are sufficient to generate an initial equilibrium
  distribution of heavy Majorana neutrinos.
  
  Baryogenesis implies stringent constraints on the light neutrino
  masses.  Assuming a similar pattern of mixings and masses for
  neutrinos and up-type quarks, as suggested by SO(10) unification,
  the observed asymmetry is obtained without any fine tuning. The
  $\n_{\m}$ mass is predicted in a range consistent with the MSW
  solution of the solar neutrino problem. $B-L$ is broken at the
  unification scale. The baryogenesis scale is given by the mass of
  the lightest of the heavy Majorana neutrinos, which is much lower
  and consistent with constraints from inflation and the gravitino
  abundance.
  
  As our discussion illustrates, the cosmological baryon asymmetry is
  closely related to neutrino properties. Already the existence of a
  baryon asymmetry is a strong argument for lepton number violation
  and Majorana neutrino masses. Together with further information
  about neutrino properties from high-energy physics and astrophysics,
  the theory of the baryon asymmetry will give us new insights into
  physics beyond the standard model.

\end{document}